\documentclass[aps,prl,twocolumn,floats,amsmath,amssymb,showpacs,floatfix]%
{revtex4}

\usepackage{graphicx}
\usepackage{dcolumn}

\begin{document}

\title{Stick-breaking model for variable-range hopping}
\author{M. Wilkinson$^{1}$, B. Mehlig$^{2}$ and V. Bezuglyy$^{1}$}
\affiliation{
$^{1}$Department of Mathematics, The
Open
University, Walton Hall,
Milton Keynes, MK7 6AA, England \\$^{2}$Department of Physics, G\"oteborg University, 41296
Gothenburg, Sweden \\}

\begin{abstract}
We consider the optimal conduction path of the one-dimensional
variable-range hopping problem. We describe a hierarchical
procedure for constructing the path which is in excellent
agreement with numerical results obtained from a percolation
approach. The advantage of the hierarchical construction is that
it is easier to analyse. We show that the distribution of hopping
lengths is well approximated by a model for the repeated breaking
of a stick at its weakest point, until the fragments are too
strong to be broken.
\end{abstract}
%
\maketitle
{\sl 1. Introduction}. The concept of variable-range hopping was
introduced by Mott \cite{Mot68} to explain the empirically observed
temperature dependence of the electrical conductivity in
disordered semiconductors at very low temperatures. Mott argued
that the conduction is determined by a competition between
transitions with large matrix elements and transitions requiring a
small activation energy: large matrix elements favour short-range
hopping, but there are unlikely to be energetically favoured
transitions at short ranges. Maximisation of the transition rate
suggests that the conductivity should have a
temperature dependence of the form \cite{Mot68}
\begin{equation}
\label{eq: 1} \sigma(T)\sim A {\rm e}^{-(T_0/T)^{1/(d+1)}}\,.
\end{equation}
It is very difficult to produce accurate quantitative results from
Mott's heuristic arguments. Ambegoakar, Halperin and Langer
\cite{Amb72} pointed out that the resistance of a sample is
determined predominantly by the highest resistance of the \lq
optimal conduction path', and demonstrated how Mott's picture is
related to percolation. They showed that the conductance is of the
form of (\ref{eq: 1}) for $d\!>\!1$, and could relate $T_0$ to the
geometry of percolating conducting clusters, which can be
determined numerically. Later Kurkij\"a{}rvi \cite{Kur73}
pointed out that in one dimension the mean value of the resistance
is determined by activation over the most unfavourable paths, and
that for (one-dimensional) wires, the mean resistance is of
Arrhenius form at very low temperatures. His argument was
improved upon by Raikh and Ruzin \cite{Rai89} who calculated the
exponential term in the mean resistance accurately. Thus the only
precise results concern the weakest link in the conduction path,
despite the fact that this problem has been investigated for
four decades. It is desirable to gain a more complete picture
of the conduction path. In this paper we consider a quantitative
model for the distribution of hopping lengths in one dimension.

Efros and Shklovskii \cite{Efr75} have suggested that electronic
correlation effects could modify the predictions of Mott's
variable-range hopping picture. Mott's approach is usually
discussed in the context of a degenerate electron gas, in which
case correlation effects may be significant. However we consider a
slightly modified problem applicable to non-degenerate systems,
thus avoiding possible complications due to correlation effects.

\begin{figure}[t]
\includegraphics[width=6cm,clip]{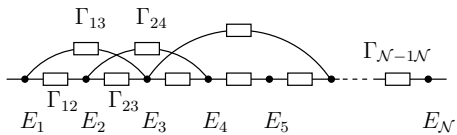}
\caption{\label{fig: 1} One-dimensional resistor-network model for
the one-dimensional variable-range hopping problem described by equation (\ref{eq: 2}).
The conductances
$\Gamma_{mn}$ are given by (\ref{eq: 3}), $E_n$ are random on-site
energies.}
\end{figure}

{\sl 2. Results.} In the limit $T\to 0 $ the one-dimensional,
non-degenerate variable-range hopping model is mapped to a
resistor-network model (Fig. \ref{fig: 1}) with conductances
$\Gamma_{mn}$. As  $T\to 0$,  conduction is confined to a single
most favourable path. Our aim is to determine the properties of
this \lq optimal conduction path'. We characterise this path by
the distribution of its hop lengths.

\begin{figure}[t]
\includegraphics[width=8cm,clip]{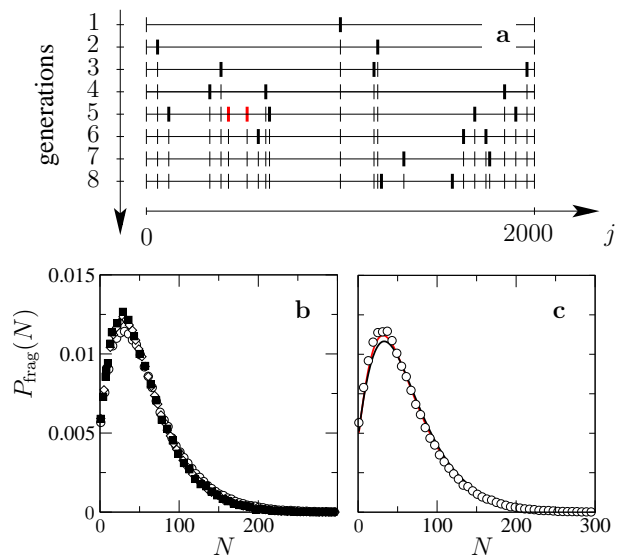}
\caption{\label{fig: 2} {\bf a} One realisation of the hierarchical
process described ($\alpha = 0.1$, $t=0.001$, ${\cal N}  = 2000$).
When segments break they do so mostly by one break (thick black
vertical lines). In one instance a segment breaks by two
simultaneous breaks (red lines) into three pieces. The process
terminates in generation $8$ when all segments are unbreakable.
For the particular realisation shown, the percolation approach
gives exactly the same segments. {\bf b} Distribution of hop
lengths of the optimal conduction path ($\blacksquare$) determined from a
percolation approach. Also shown are approximations employing the
hierarchical process, allowing for one ($\circ$) and two ($\Diamond$)
breaks per segment. Allowing for more than two simultaneous breaks
per interval does not change the distribution significantly.
The parameters were $\alpha = 0.1$, $t = 0.002$, ${\cal N} = 5000$.
{\bf c}
Stick-breaking approximation
 for $P_{\rm hop}(N)$ allowing for one break per segment
(with the same parameters as in {\bf b}). We plot equation (\ref{eq: 11}), black line,
and on two iterations of (\ref{eq: 10}), red line.
Further iterations do not change the red line noticeably.
For comparison we show data from the hierarchical scheme allowing one
break per segment ($\circ$), as in  {\bf b}.
}
\end{figure}

We introduce a hierarchical approximation scheme for the optimal
conduction path. An example is shown in Fig. \ref{fig: 2}{\bf a}.
In this example, the hierarchical construction yields exactly the
same path as a numerical percolation approach (described below).
For the parameters in Fig. \ref{fig: 2}{\bf a}, the hierarchical
construction yields the correct hops in $98.2\%$ of all cases. The
corresponding distribution of hop lengths is shown in Fig.
\ref{fig: 2}{\bf b}, in excellent  agreement with results from a
percolation approach. The advantage of the
hierarchical procedure is that it is susceptible to geometrical and
statistical analysis. We show that the
problem can be analysed in terms of repeatedly
breaking a stick at its weakest point until the fragments cannot be broken down further. The
distribution of fragment lengths is found to be in
very good agreement with the distribution of hopping lengths
(Fig. \ref{fig: 2}{\bf c}).

By microscopic probes (such as those used in tunneling
microscopy) it is becoming possible to determine the paths
involved in carrying the current in microscopic systems. The
distribution of hopping lengths may thus become accessible to experiments.

The remainder of this letter is organised as follows. First we
consider a resistor-network model for the one-dimensional,
non-degenerate variable-range hopping problem.  We then describe
our percolation approach to determining the optimal conduction
path (an extension of the approach employed in \cite{Lee85}). Our
results serve as a benchmark for the hierarchical approach which
is described next. The remaining sections show how to compute
the distribution of hop lengths in terms of a stick-breaking process.

{\sl 3. Resistor-network model.} We consider an array of sites
with energies $E_n$, independent identically distributed random
variables in $[0,1]$. We assume that the occupation probabilities
$P_n$ satisfy a rate equation (neglecting quantum-mechanical
interference effects)
\begin{equation}
\label{eq: 2} \dot P_n=\sum_m R_{mn}P_m-P_n\sum_m R_{nm}\,.
\end{equation}
The transition rate from state $n$ to state $m$ is called $R_{nm}$;
the principle of detailed-balance implies
${R_{nm}={R_{mn}}}\exp[-(E_m-E_n)/(k_{\rm B}T)]$.
We specify the rates by assuming that downward
transitions ($E_m \!<\! E_n$) occur at a rate
$R_{mn}= \epsilon^{\vert n-m\vert}$
for $\epsilon>0$.

The master equation (\ref{eq: 2}) is inconvenient in its
original form, because its equilibrium distribution $
P^{(0)}_n=Z^{-1}\exp[-E_n/(k_{\rm B}T)]$ is non-uniform. We
therefore transform it to an equation evolving towards a uniform
equilibrium density, and then describe deviations from this
equilibrium in terms of equations analogous to Kirchoff's laws for
a resistor network, with effective conductances $\Gamma_{mn}$
between lattice sites $m$ and $n$. We then expect (following the
line of argument employed in \cite{Amb72}) that the diffusion rate
is determined by a resistor network (Fig. \ref{fig: 1}) with \lq
conductances' $\Gamma_{mn} = P^{(0)}_m R_{nm}$:
\begin{equation}
\label{eq: 3}
\Gamma_{mn}=
\epsilon^{\vert n-m\vert}{\rm e}^{-{\rm max}(E_m,E_n)
/(k_{\rm B}T)}
\equiv {\rm e}^{-\Delta_{mn}/t}\,.
\end{equation}
The last equality defines  \lq distances'
$ \Delta_{mn}=\alpha t \vert n-m \vert +{\rm max}(E_m,E_n)$,
with $\alpha = -\log \epsilon$, and $t = k_{\rm B} T$.

{\sl 4. Percolation approach.} In the limit as $t\to 0$, ratios
between the conductances $\Gamma_{mn}$ become very large. In this
limit the fraction of the current in any bond is almost certainly
very close to zero (\lq inactive bond'), or else very close to
unity (\lq active bond'). In \cite{Amb72} it was pointed out that
the resistance of the network is almost surely well approximated
by the resistance of the active bond with the highest
resistance. This bond can be identified by mapping to a
percolation problem: for a given choice of threshold conductance
$\Gamma_0$, bonds with conductances $\Gamma_{mn}\!<\!\Gamma_0$
are removed from the network. Decreasing $\Gamma_0$, we
test for connectivity of the network. As $\Gamma_0$ decreases
below the conductance of a \lq critical' bond, $(n,m)^\ast$
(with conductance $\Gamma_{nm}^\ast$), we find that the network
becomes connected. The conductance of the network is
expected to be well approximated by $\Gamma_{mn}^\ast$. Now we
repeat this procedure for the sub-networks to the left and to the
right of the critical bond, determining the critical bonds of the
sub-networks. Repeating this procedure results in identifying
critical bonds in progressively shorter sections of the network.
The process terminates when the critical bond for a sub-network is
the one bond that connects its ends directly. In this way the
optimal conduction path is found. This approach is accurate when
$t\rightarrow 0$, but the search over all possible paths is very
time consuming, and not susceptible to statistical analysis. Our
objective was to find a way of determining the optimal conduction
path by repeated application of simple rules -- which can be
analysed statistically.

{\sl 5. Hierarchical model.} Our approach consists of breaking
down the network by proposing a succession of current-carrying
paths with smaller and smaller resistances. At each stage we
consider a segment $(m,n)$ of the network, between nodes $m$ and
$n$. We ask whether the conductance of the direct connection
$\Gamma_{mn}$ is less than that of an indirect path, which goes
through an intermediate node $j$. If the resistance
$\Gamma^{-1}_{mj}+\Gamma^{-1}_{jn}$ is less than
$\Gamma^{-1}_{mn}$, then the network is broken at node $j$, and we
consider the sub-networks $(m,j)$ and $(j,n)$.
If there is more than one
break which lowers the resistance, we choose the one which gives
the lowest resistance. If there is no single \lq break' lowering
the resistance, we try inserting two simultaneous breaks, then
three, etc. If there is more than one possible way to insert the
breaks, the choice giving the lowest resistance is used. If the
resistance cannot be reduced below that of the direct connection
by considering any sequence of nodes connected in series, then the
sub-network is determined to be \lq unbreakable'.

There is no guarantee that the process will produce the optimal
conduction path, only that the resistance will decrease with each
refinement. As Fig. \ref{fig: 2} shows, however, this procedure
provides an excellent approximation to the optimal conduction path.

{\sl 6. Stick-breaking model.} The hierarchical method has the
advantage that it is easier to analyse than the percolation
approach. In the remainder
we illustrate this by
showing that the simplest hierarchical model (allowing only
one break per segment at a time) can be approximated by a \lq
stick-breaking model'.

It is found that when an interval can be broken at one point to
reduce the resistance, the break point usually corresponds to the
site within the interval with the smallest energy $E_j$. This observation leads to the
following simplified model for the hierarchical process. When
considering an interval between sites $n$ and $m$ (with $n>m$), we
determine whether the resistance is reduced by inserting a single break.
If so, we determine the  smallest energy $E_j$ with $m<j<n$, and break the
interval there. The process of
sub-division is continued until all of the intervals are
determined to be unbreakable.

In order to model this process we introduce two probabilities.
Let $P_{\rm br}(N)$ be the probability that a fragment of length $N$
elements is breakable. If it is breakable, it will break at a position
$M$ elements from the end with probability $p(M,N)$. Starting from
and infinitely long stick and iterating this process, we end up with
a set of unbreakable fragments with a
probability $P_{\rm frag}(N)$ for having length $N$. In the
following, we discuss how $P_{\rm br}(N)$ and $p(M,N)$ respectively are chosen
to make this stick-breaking model correspond to the simplified
hierarchical model of variable range hopping. We then discuss how
the stick-breaking model is solved. Figure \ref{fig: 2}{\bf c} shows
that the resulting distribution
of fragment lengths, $P_{\rm frag}(N)$, is a good approximation to
the distribution of hopping lengths, $P_{\rm hop}(N)$.

\begin{figure}[t]
\includegraphics[width=6.6cm,clip]{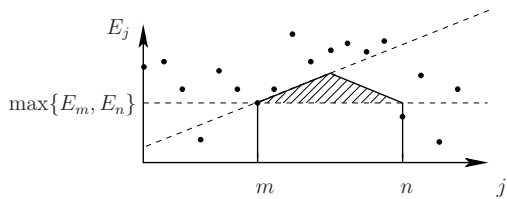}
\caption{\label{fig: 3} Shows points in $(j,E_j)$ plane ($j =
0,1,2,\ldots$ and $E_j$ uniformly distributed on $[0,1]$). The
probability that the segment $(m,n)$ is unbreakable (with a single
break) is equal to the probability that the \lq roof of the house'
(hashed) of area $A'$ is empty. The dashed horizontal line
represents $\max\{E_m,E_n\}$, the second dashed line $E_j =
\max\{E_m,E_n\}+\alpha t\,(j-m)$.}
\end{figure}

{\sl 7. Probability for a segment to break.} The set of energies
can be represented by a set of points in the $j,E_j$ plane, which
are randomly distributed in energy, with the density of points in
the plane being unity.

 In the limit of
 $t\to 0$, the resistance of an indirect path through one intermediate node is approximated by
 the larger resistance of the two links, so that we write
 $\Gamma^{-1}_{mj}+\Gamma^{-1}_{jn} \sim \exp [{\rm max}(\Delta_{mj},\Delta_{jn})/t]$.
So the probability for a segment $[m,n]$ to be
unbreakable is the probability that $\Delta_{nj}>\Delta_{nm}$ and
$\Delta_{jm}>\Delta_{nm}$ for any choice of $j$ (with $m<j<n$).
These two conditions imply that the regions below two lines in the
$(j,E_j)$ plane are empty, in the interval $[m,n]$. Thus the
probability that the interval $[m,n]$ is unbreakable (with a
single break) is equal to the probability of finding the \lq
house' in Fig. \ref{fig: 3} empty.

The probability of finding a region in the $(j,E_j)$ plane of area
$A$ to be empty of points is $\exp(-A)$. At first sight it seems
as if we should take $A$ to be the area of the \lq house' in
figure \ref{fig: 3} and the probability that a segment is breakable
would be $P_{\rm br} = 1-{\rm e}^{-A}$. We do however have some
prior knowledge about the energies: because we always break at the
minimum energy, the none of the energies $E_j$ in the interval
$[m,n]$ is less than ${\rm max}(E_n,E_m)$. Thus it is the area of
the \lq roof' of the house (hashed in Fig. \ref{fig: 3}) which is
relevant, and the probability for this segment being breakable is
therefore
\begin{equation}
\label{eq: 4} P_{\rm br}(n-m) =1-\exp(-A')\ ,\ \ \
A'=\alpha t(n-m)^2/4\,.
\end{equation}
Thus the probability for a
segment being breakable in our construction is, in fact, solely
dependent on its length, and given by (\ref{eq: 4}).

{\sl 8. Distribution of break points.} Next we consider the
probability $p(M,N)$ that an interval of length $N$ will break at
a position $M$ units from one of its ends. Because the energies $E_j$ are
independently and identically distributed, we might expect that
the position of minimal $E_j$ is uniformaly distributed, that is
$p(M,N)$ should be independent of $M$. This is incorrect, however,
because we are considering the distribution of energies in an
interval which is selected to be breakable.

The probability  $p(M,N)$ for the position $M$ of minimal $E_j$ in
an interval of length $N$ is symmetric around the midpoint of the segment. Let
$x\in [0,1]$ be a variable measuring distance of the break from the end of the
interval, setting $x=2M/(N-1)$ if $M<N/2$ and $x=2(N-M)/(N-1)$ if $N/2<M<N$. We
consider the continuous probability density $p(x)$ for $x$, which is related to
$p(M,N)$ by:
\begin{equation}
\label{eq: 5} p(M,N) = {\frac{1}{N-1}}\left\{
\begin{array}{ll}
p\big({2 M\over N\!-\!1}\big) & \mbox{if $0< M< N/2$}\\[1mm]
p\big({2N-2M\over N\!-\!1}\big) & \mbox{$ N/2 \leq M < N$.}
\end{array}
\right .
\end{equation}
The density $p(x)$ is calculated by geometrical analysis
of Fig. \ref{fig: 3}.  We rescale the energy $E_j$
to a new variable $y=(N-1)[E_j-{\rm max}(E_n,E_m)]$ so
that the density of points in the $(x,y)$
plane is unity.
In these new coordinates the line determining
the boundary of the region which
must be occupied to ensure breakability is $y=\nu x$
with $\nu =\alpha t  N^2/2$.
Instead of calculating $p(x)$ we compute the complementary
probability density $q(x)$, that is
the probability density for the position of the minimum energy in an
unbreakable segment. The probability density for the location
of minima irrespective of
breakability is uniform (and by normalisation equal to unity), so
that $1=P_{\rm br}p(x)+(1-P_{\rm br})q(x)$, or:
\begin{equation}
\label{eq: 6}
p(x)=[{1-\exp(-\nu/2)q(x)}]/[{1-\exp(-\nu/2)}]\,.
\end{equation}
It is easier to compute $q(x)$ because the conditional knowledge
about unbreakable (at one break) intervals is simpler. These
intervals have no points in the $(x,y)$ plane inside the \lq forbidden'
triangle $0<y\!<\!\nu x$, $0\!<\!x\!<\!1$. Thus $q(x)$ is the distribution of
$x$ for the point with minimal $y$ from a random, unit density
scatter in the \lq allowed' region $y\!>\!\nu x$, $0\!<\!x\!<\!1$.
Let $F(y_0)$ be the
probability that the point with lowest value of $y$ is above $y_0$.
This satisfies $F(0)=1$ and
${\rm d }F/{\rm d}y=-{\rm d} A/{\rm d}y\,F$
where $A(y)$ is the area of the allowed region below $y$. We have
\begin{equation}
\label{eq: 7}
q(x)=\int_0^\infty \!\!\!\!{\rm d}y\, \bigg\vert \frac {dF}{dy}\bigg\vert
\,\chi_{[0,x(y)]}(x)\,\frac{1}{x(y)}
\end{equation}
where $\chi_{[a,b]}(x)$ is the characteristic function (equal to unity on the interval
$[a,b]$ and zero elsewhere) and $x(y)={\rm d}A/{\rm d}y$ is the width of the allowed region
at $y$. Using the differential equation for $F(y)$ we obtain
$q(x)=\int_{\nu x}^\infty {\rm d}y\ F(y)$. We find $F(y)={\rm e}^{-y^2/2\nu}$ for
$0<y<\nu$ and $F(y)={\rm e}^{-y+\nu/2}$ for $y>\nu$ and finally obtain:
\begin{eqnarray}
\label{eq: 8}
q(x)
&=&{\rm e}^{-{\nu}/{2}}\!+\!\sqrt{{\pi \nu\over{2}}}\biggl[{\rm
erf}\biggl(\sqrt{{\nu\over 2}}\biggr)\!-\!{\rm
erf}\biggl(x\sqrt{\nu\over 2}\biggr)\biggr] \,.
\end{eqnarray}
Eqs. (\ref{eq: 5}-\ref{eq: 8}) show that segments are more likely
to break in the centre, as expected.  This bias is the stronger
the shorter the segments are. For long
segments ($\nu\to\infty$) we obtain $p(M,N) \approx 1/(N-1)$.
We find that a good approximation to $q(x)$ is
\begin{equation}
\label{eq: 9}
q(x) \approx 1  +\nu\,(1/2-x)\,.
\end{equation}

{\sl 9. Self-consistent solution of the stick-breaking process}.
We now determine the probability distribution $P_{\rm frag}(N)$ of
the lengths of the unbreakable fragments, by extending a calculation
for the distribution of strength of a repeatedly broken random chain,
discussed in \cite{Wil07}. We start from a very
long stick, and after $T$ steps of this process, we have $W_{\rm
D}(N,T)$ segments of length $N$ which have been determined to be
unbreakable, and $W_{\rm U}(N,T)$ segments of length $N$ which
have not yet been tested. Assuming that these numbers are so large
that it is sufficient to consider expectation values,  we have the
recursion $ W_{\rm U}(N,T+1)=2\sum_{M=N+1}^\infty p(N,M) P_{\rm
br}(M) W_{\rm U}(M,T)$ and $ W_{\rm D}(N,T+1)=[1-P_{\rm
br}(N)]\,W_{\rm U}(N,T)$. Rather than following this iteration for
a single stick being broken, we consider a steady state of $W_{\rm
U}(N,T)$, with destruction of one additional stick being initiated
at each step:
\begin{equation}
\label{eq: 10}
W(N)=2\sum_{M=N+1}^\infty p(N,M) P_{\rm br}(M) W(M)\,.
\end{equation}
The corresponding probability distribution of fragment
lengths (or hop lengths) is
$ P_{\rm frag}(N)=Z^{-1}[1-P_{\rm br}(N)]W(N)$
where $Z$ is a normalisation factor.

When $p(N,M) = 1/(M-1)$, equation (\ref{eq: 10}) is solvable.
Replacing sums by integrals, we obtain
$W_0(N)\propto N^{-2}\, {\rm e}^{-{\rm Ei}(\alpha tN^2/4)}$,
where ${\rm Ei}$ is the exponential integral. When $p(M,N)$ is
given by eqs. (\ref{eq: 5}), (\ref{eq: 6}), and (\ref{eq: 8}), we solve (\ref{eq: 10})
by iteration starting with $W_0$. Usually a few iterations give an accurate solution. The first iteration
gives [using eq. (\ref{eq: 9}) to approximate (\ref{eq: 8})]
\begin{eqnarray}
\label{eq: 11}
P_{\rm frag}(N) &\propto& {\rm e}^{-\nu/2}[I_1(\nu)+I_2(\nu)]\\
I_1(\nu) \!&=&\! \int_\nu^{4\nu}
\frac{{\rm d}\mu}{\mu^2}
\big({\rm e}^{\mu/2}\!-\!1\!+\!{3\mu\over 2}\!-\!2\sqrt{\mu\nu}\big)
{\rm e}^{-{\rm Ei}(\mu/2)-\mu/2}\nonumber\\
I_2(\nu) \!&=&\! \int_{4\nu}^\infty \frac{{\rm d}\mu}{\mu^2}
\big({\rm e}^{\mu/2}\!-\!1\!-\!{\mu\over
2}\!+\!2\sqrt{\mu\nu}\big) {\rm e}^{-{\rm
Ei}(\mu/2)-\mu/2}\nonumber\,.
\end{eqnarray}
The distribution $P_{\rm frag}(N)$ given by (\ref{eq: 11}) is shown
in Fig. \ref{fig: 2}{\bf c}. Results obtained
from higher iterations of (\ref{eq: 10}) using (\ref{eq: 8}) are also
shown. They converge rapidly and provide a strikingly
accurate approximation to $P_{\rm hop}(N)$.

Support from Vetenskapsr\aa{}det is gratefully
acknowledged.


\begin{thebibliography}{0}

\bibitem{Mot68}  N. F. Mott, {\sl J. Non-Cryst. Solids}, {\bf 1}, 1,
(1968).

\bibitem{Amb72} V. Ambegoakar, L. S. Langer and B. I. Halperin, {\sl
Phys. Rev. B},   {\bf 4}, 2612 (1971).

\bibitem{Kur73} J. Kurkij\" arvi, {\sl Phys. Rev. B}, {\bf 8},
622, (1973).

\bibitem{Rai89} M. E. Raikh and I. M. Ruzin, {\sl Sov. Phys.
JETP}, {\bf 66}, 642-7, (1989)\ [{\sl Zh. Eksp. Teor. Fiz.}, {\bf
95}, 1113-22, (1989).]

\bibitem{Efr75} A. L. Efros and B. I. Shklovskii,
{\sl J. Phys. C: Solid State Phys. }, {\bf 8}, L49, 1975

\bibitem{Lee85} P. A. Lee, {\it Phys. Rev. Lett.}, {\bf 53}, 2042,
(1984).

\bibitem{Wil07} M. Wilkinson and B. Mehlig, {\it J. Stat. Phys.},
{\bf 127}, 1279, (2007).

\end{thebibliography}
\end{document}